\documentclass[12pt]{article}
\usepackage{aas_macros}
\usepackage{fullpage}
\usepackage{graphicx}
\usepackage[suffix=]{epstopdf}
\usepackage{natbib}
\usepackage{amsmath}
\usepackage{url}
\usepackage{xspace}
\usepackage{color}
\usepackage{deluxetable}

\def\lesssim{\mathrel{\hbox{\rlap{\hbox{\lower3pt\hbox{$\sim$}}}\hbox{\raise2pt\hbox{$<$}}}}}
\def\gtrsim{\mathrel{\hbox{\rlap{\hbox{\lower3pt\hbox{$\sim$}}}\hbox{\raise2pt\hbox{$>$}}}}}

\begin{document}

\title{{\large Maximizing Science in the Era of LSST:\\ A Community-based Study of Needed US OIR Capabilities}\\\vspace{0.5in}
Stars Study Group Report:\\ Rotation and Magnetic Activity in the Galactic Field \\ Population and in Open Star Clusters}
\author{
Suzanne L. Hawley (University of Washington), \\
Ruth Angus (University of Oxford), \\
Derek Buzasi (Florida Gulf Coast University), \\
James R. A. Davenport (Western Washington University), \\
Mark Giampapa (National Solar Observatory),\\
Vinay Kashyap (Harvard-Smithsonian Center for Astrophysics), \\
S\o ren Meibom (Harvard-Smithsonian Center for Astrophysics)
\vspace{0.1in}  \\
{July 13, 2016}
}

\date{} 

\maketitle

\section{Introduction and Background}

The Large Synoptic Survey Telescope (LSST) will provide precise ground-based
photometric monitoring of billions of stars in the Galactic field and in open star clusters.
The light curves of these stars will give an unprecedented view of the evolution of
rotation and magnetic activity in cool, low-mass main-sequence dwarfs of spectral type GKM, allowing precise calibration of rotation-age and flare rate-age relationships, and opening a new window on the accurate age dating of stars in the Galaxy.
Previous surveys have been hampered by small sample size, poor photometric precision and/or short time baselines, so LSST data are essential for obtaining new, robust age calibrations.

The evolution of the rotation rate and magnetic activity in solar-type stars are intimately connected.
Stellar rotation drives a magnetic dynamo, producing a surface magnetic field and magnetic activity which manifests as starspots, chromospheric (Ca II HK, H$\alpha$) and coronal (X-ray) emission and flares.  The magnetic field also drives a stellar wind causing angular momentum loss  (``magnetic braking'') which slows the rotation rate over time, leading to decreased magnetic activity.
More magnetically active stars (larger spots, stronger Ca II HK, H$\alpha$ and X-ray emission, more flares) therefore tend to be younger and to rotate faster.  The rotation-age relationship is known as gyrochronology, and the correlation between rotation, age and magnetic activity for solar-type stars was first codified by \citet{skumanich1972}.
However, the decrease in rotation rate and magnetic field
strength over long time-scales is poorly understood and, in some cases, hotly
contested \citep[{e.g.}][]{angus2015, van-saders2016}.
Recent asteroseismic data from the Kepler spacecraft have revealed that
magnetic braking may cease at around solar Rossby number, implying that
gyrochronology is not applicable to older stars \citep{van-saders2016}.

In addition, the rotational behavior of lower mass stars is largely unknown due
to the faintness of mid-late type M dwarfs.  There is reason to believe that M dwarfs cooler than spectral type $\sim$ M4 may behave differently from the G, K and early M stars, since that spectral type marks the boundary where the star becomes fully convective, and a solar-type shell dynamo (which requires an interface region between the convective envelope and radiative core of the star) can no longer operate.  Using chromospheric H$\alpha$ emission as a proxy, \citet{west2008} studied a large sample of M dwarfs from SDSS and showed that magnetic activity in mid-late M dwarfs lasts much longer than in the earlier type stars.  

LSST will provide photometric rotation periods for a new region of
period-mass-age parameter space.
The Kepler spacecraft focused on Earth-like planets with Sun-like hosts,
thus the majority of its targets were G type, with fewer K and M dwarfs.
Unlike Kepler however, any target falling within LSST's field of view will
be observed --- not just those on a predetermined target list.
In addition, due to the large collecting area of LSST, it will be sensitive
to a significant population of distant K and M dwarfs.
 LSST will operate for 10 years, more than double the length of
the Kepler prime mission.
This long time baseline will enable rotation signatures of faint, slowly rotating
stars to be detected, populating both low-mass and old regions of the
age-rotation parameter space.
Thus, LSST will provide an important complementary data set to
Kepler (and the upcoming TESS mission).  

The LSST data will also allow an unprecedented view of stellar flares, and the calibration of flare rate-age relations that may provide an additional method for age-dating M dwarf populations.  \citet{kowalski2009} used sparsely sampled SDSS light curves in Stripe 82 to quantify M dwarf flare rates as a function of height above the Galactic plane, and showed that flare stars may comprise a younger population than active stars (those showing H$\alpha$ emission).  Long time baselines and monitoring of large numbers of stars are required to obtain good flare statistics, so LSST will be perfectly suited for this study.

As coeval, equidistant, and chemically homogeneous collections of stars, open star clusters with different ages are ideal for studying the dependencies of astrophysical phenomena on the most fundamental stellar parameters - age and mass. Indeed, there are few fields in astronomy that do not rely on results from cluster studies, and clusters play a central role in establishing how stellar rotation and magnetic activity can be used to constrain the ages of stars and stellar populations.  In particular,
clusters provide essential calibration for rotation-age-activity relations, since each cluster gives a snapshot of stellar evolution at a single age, for all masses. 
\citep[e.g.][]{meibom2009,meibom2011,mms+11,mbp+15,ghr+06,gondoin2012,gondoin2013,wdm+11}.

LSST will also enable the use of cluster and field stars as  laboratories for investigating magnetic activity cycles (such as the 11 year cycle on the Sun).  There is some evidence that younger, more active stars are less likely to show regular cycle behavior, while older stars such as the Sun typically do show regular cycles \citep{baliunas1995}.  Due to the long monitoring times that are required to diagnose activity cycles, it has previously been difficult to carry out a large-scale survey of activity cycle behavior, and thus quantify the changes that occur with magnetic dynamo evolution (e.g. as the star spins down).  LSST will easily rectify this situation and indeed the cadence will be well-suited to observing cyclic behavior both in the field and in clusters.  The changes that occur in the surface magnetic field (both strength and topology) as a star ages are also not well understood, with fewer than 100 (nearby, bright) stars presently having good measurements.  Followup observations of stars from a large LSST sample covering a range of ages and masses with well-determined cycle periods will open a new window on the study of magnetic field evolution.

While LSST light curves have the potential to answer some of the most fundamental
questions regarding the evolution of stellar rotation and magnetism, it is
essential that the properties of the target stars be accurately determined.  In order to understand the scope of the followup observations that will be needed to characterize magnetically active stars that exhibit starspots and flares, we performed a number of simulations.  We first describe simulations of field stars at several galactic latitudes, sampled with an LSST cadence, and examine the target densities of stars with detected rotation periods (due to starspot modulation) and detected flares.  We then consider open clusters at different ages, predict rotation and flare rates, and discuss the complications of determining cluster membership.  Finally we look at the constraints for determining activity cycles and measuring magnetic fields directly.  Using the results of these simulations, we outline the followup requirements necessary to fully
exploit the LSST data set for stellar rotation and magnetism studies both in the Galactic field and in open clusters.

Although we focused our study here on magnetically active stars, in order to provide well-defined estimates for followup capabilities, we note that similar observing strategies and followup resources will be valuable for investigations of a wide variety of variable stars, including eclipsing binaries, pulsating stars at a wide range of periods from RR Lyrae to Cepheids to LPVs, cataclysmic variables and novae, and also for other Galactic variability phenomena such as microlensing and planetary transits.   Wide field followup imaging and spectroscopic (both moderate and high resolution) facilities will be useful for all of these stellar science topics.

\section{Simulating LSST field star samples}

\subsection{Cadence model}

We used the $minion\_1016$ cadence model, which is the most recent (May 2016) ``baseline cadence'' being tested for LSST simulations\footnote{\url{https://github.com/LSSTScienceCollaborations/ObservingStrategy}}.  This model has some deviations from a regular, every few days, cadence, including fewer u and g band observations, and very few observations at high galactic latitude (e.g. b = -80).  However, we have adopted it for the purposes of this study, and have only considered galactic latitudes between -60 and -10 degrees.

\subsection{Field Star Populations}

We used the TRILEGAL \citep{trilegal} galaxy simulation code to generate
field stellar populations for hypothetical LSST fields.
These fields were centered on the same galactic longitude, $l=45$ degrees, and four
different galactic latitudes:  $b=-10, -20,-40,-60$ degrees.  Each TRILEGAL field
comprises a catalogue of stars with properties including age,
effective temperature, gravity, and $ugriz$ magnitudes.

\begin{figure}[]
\centering
\includegraphics[width=3in]{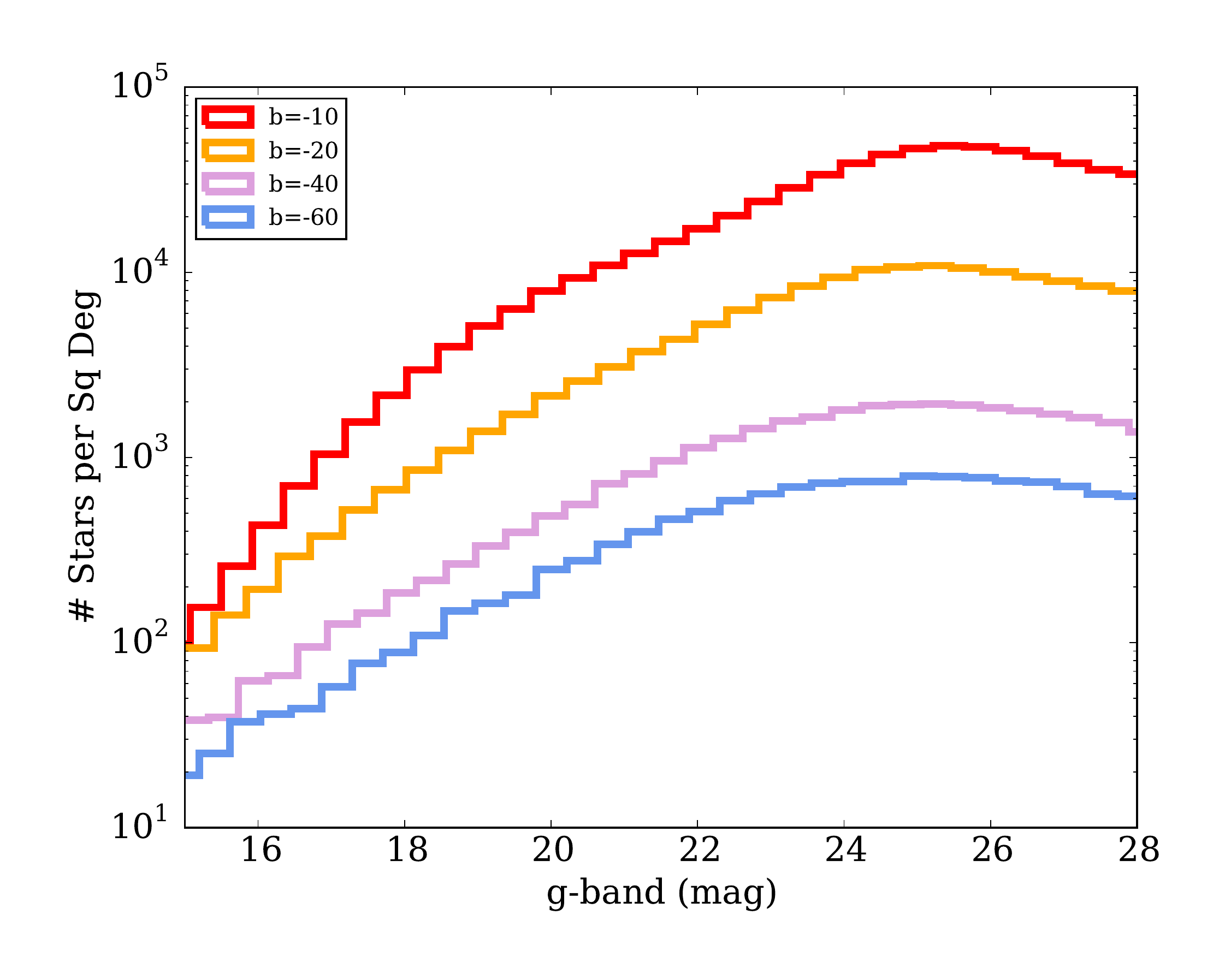}
\includegraphics[width=3in]{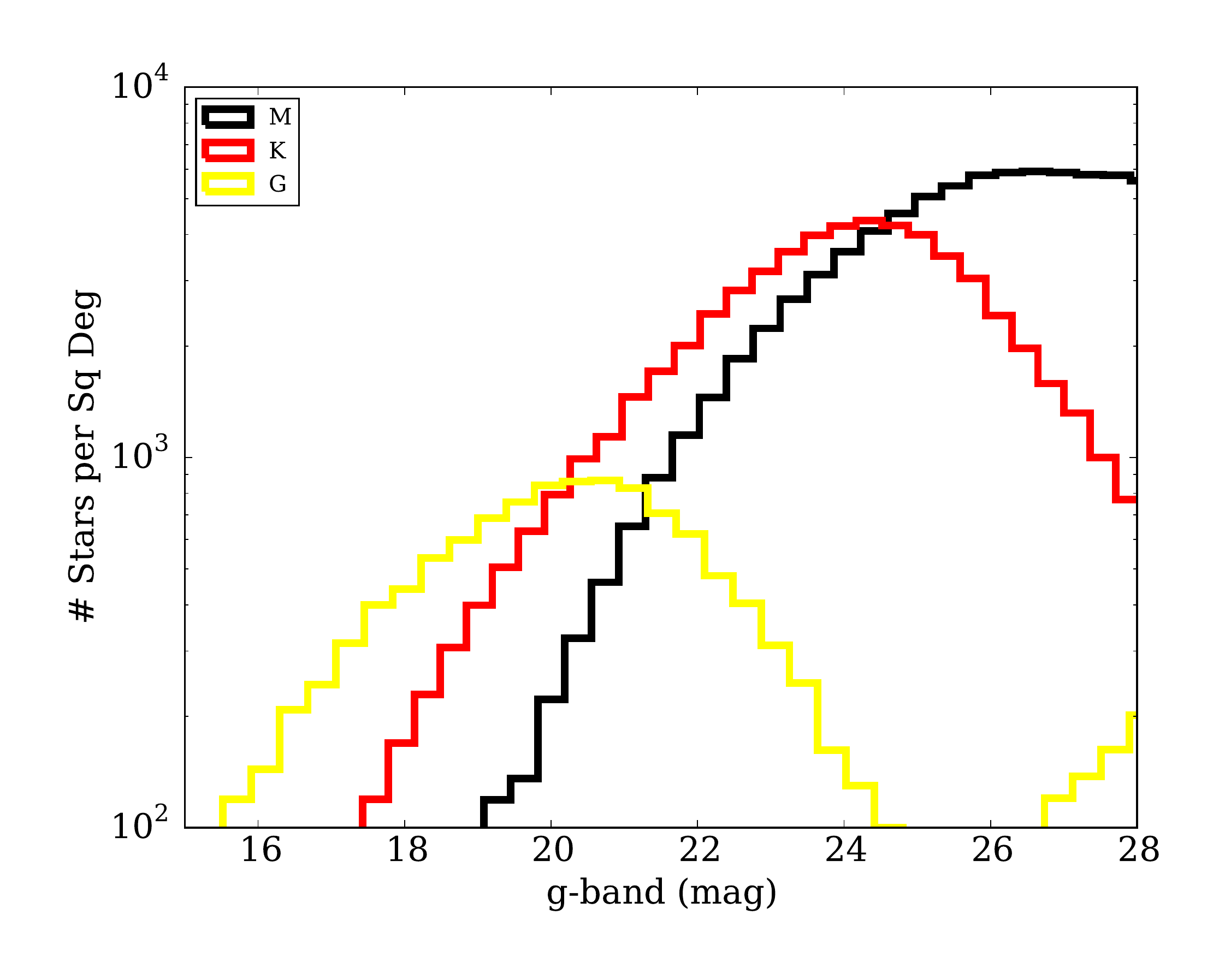}
\caption{
Left: Density of stars along each of the lines of sight as a function of apparent magnitude. Right: Density of G, K, and M dwarfs in the $b=-20$ TRILEGAL field.}
\label{fig:fieldsim}
\end{figure}

Figure \ref{fig:fieldsim} shows the target distributions for the four fields as a function of apparent magnitude.  In total, there are nearly a
million stars per square degree in the lowest latitude ($b$ = -10 degrees)
field, dropping to only a few thousand stars per square degree in the field at the highest
latitude studied ($b$ = -60 degrees).  The right panel shows the distributions of G, K and M dwarfs for the representative field at $b$ = -20 degrees.  There are about 10 times as many M dwarfs ($3000<T<4000$K)
and 6 times as many K dwarfs ($4000<T<5000$K) as there are G dwarfs ($5000<T<6000$K) in these fields, but the G dwarfs are much brighter (averaging around g=20) while most of the M dwarfs have g$>$ 24.

We randomly selected stars from each field in order to produce
representative but manageable target samples to investigate for rotation periods and flares.  Selected stars had $r$-band magnitudes between 16 and 28 (because of the LSST
magnitude limits) and $\log g$ $>4$ (because we want to select main sequence
dwarfs).

\subsection{Rotation Simulations}

Twenty thousand stars were randomly chosen in each field, and rotation periods
were calculated using the \citet{angus2015} gyrochronology relation, which
converts ages and $B-V$ color (calculated from TRILEGAL\ $g-r$) into rotation
periods. 
Converting $B-V$ color to $g-r$ was done using the transformations from Table 1 of \citet{jester2005}. 
Code similar to that used in \citet{aigrain2015} was used to
simulate light curves by placing dark starspots on a rotating sphere
and integrating the total resulting flux over the surface.
Stellar flux variations produced by dark active regions on the surface are
typically non-sinusoidal and this starspot model provides a more accurate
representation of stellar light curves than a simple sinusoid.
However, for simplicity we fixed the mean starspot lifetime at 30.5 days for all simulations
and did not include differential rotation.  Both rapid starspot evolution and differential rotation will result
in quasi-periodic light curves that will be somewhat more difficult to recover, so our results should be
considered `best case'.

In order to assign appropriate amplitudes to the simulated light curves we
approximated the relation between rotation period, amplitude of variability
and $T_{eff}$, based on the \citet{mcquillan2014} Kepler sample.
We then assigned amplitudes by drawing values from Gaussians with means
corresponding to the mean amplitudes of stars with similar $T_{eff}$ and
$P_{rot}$ in \citet{mcquillan2014} and variances given by the variance in each
bin.  White noise was added to the light curves, with amplitudes that depended on
the $r$-band magnitude, based on the values provided in \citet{lsst,jacklin2015}\footnote{More detailed
simulations would include systematic and/or correlated noise sources as determined for real LSST data, however these are not yet available.  Systematic effects will limit robust period recovery, again making our results somewhat optimistic.}.
We sampled these light curves using the LSST cadence model described above, and attempted to
recover their rotation periods using a Lomb-Scargle (LS) periodogram
\citep[][]{lomb1976, scargle1982}.
LS periodograms were computed for each light curve over a grid of 1000 periods
ranging from 2 to 100 days.
The position of the highest peak in the periodogram was recorded as the
measured period.

\begin{figure}
\begin{center}
\includegraphics[width=3.5in, clip=true]{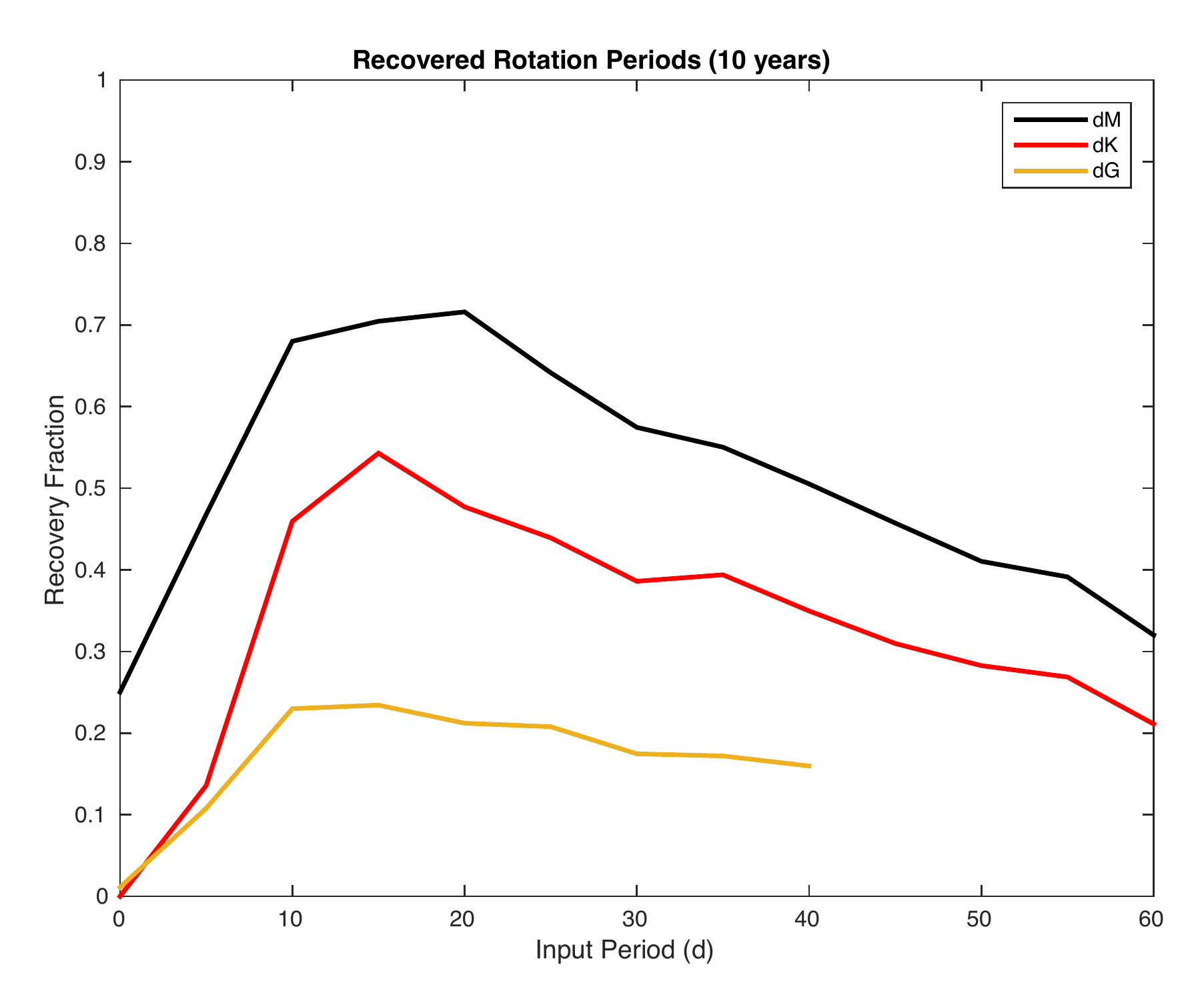}
\caption{The fraction of recovered rotation periods as a function of input
rotation period for G, K and M dwarfs.
Relatively few rotation periods shorter than around ten days are recovered for
all spectral types.
This is likely due to the three day minimum LSST observing interval.
Rotation period sensitivity drops off at longer periods since more slowly
rotating stars have smaller variability amplitudes.
There are no G stars with rotation periods greater than $\sim$ 40 days, since
these stars would be older than the Galaxy.
The rotation period sensitivity of LSST peaks at $\sim$ 20 days.}
\label{fig:period_recovery_fractions}
\end{center}
\end{figure}

We computed the recovery rates of injected periods for 1, 5 and 10 years of LSST monitoring data, and show these rates as a function of period in Figure \ref{fig:period_recovery_fractions} for each spectral type, in the $b$= -20 field. Successful period recovery was defined by output periods within 10\% of the input values.
Figure \ref{fig:period_recovery_fractions} illustrates the rotation period
sensitivity of LSST which peaks near 20 days, dropping towards shorter periods due
to the relatively large ($\sim$ three day) interval between observations.
The sensitivity also drops towards long periods due to the smaller variability
amplitudes (less starspots) for slowly rotating stars.  Note that the model gyrochronology relations do not predict G stars with periods longer than about 40 days at the age of the Galaxy.

Using the full ten-year data set we are able to accurately recover 60-70\% of
rotation periods for stars with $T_{\mathrm{eff}}>4500$ K and 70-80\% for
stars with $T_{\mathrm{eff}}<4500$ K, brighter than 23rd magnitude.  As discussed above these rates are probably optimistic, but are likely within a factor of two of the actual recovery rates once all the mitigating factors are included.

\begin{figure}[]
\centering
\includegraphics[width=3in]{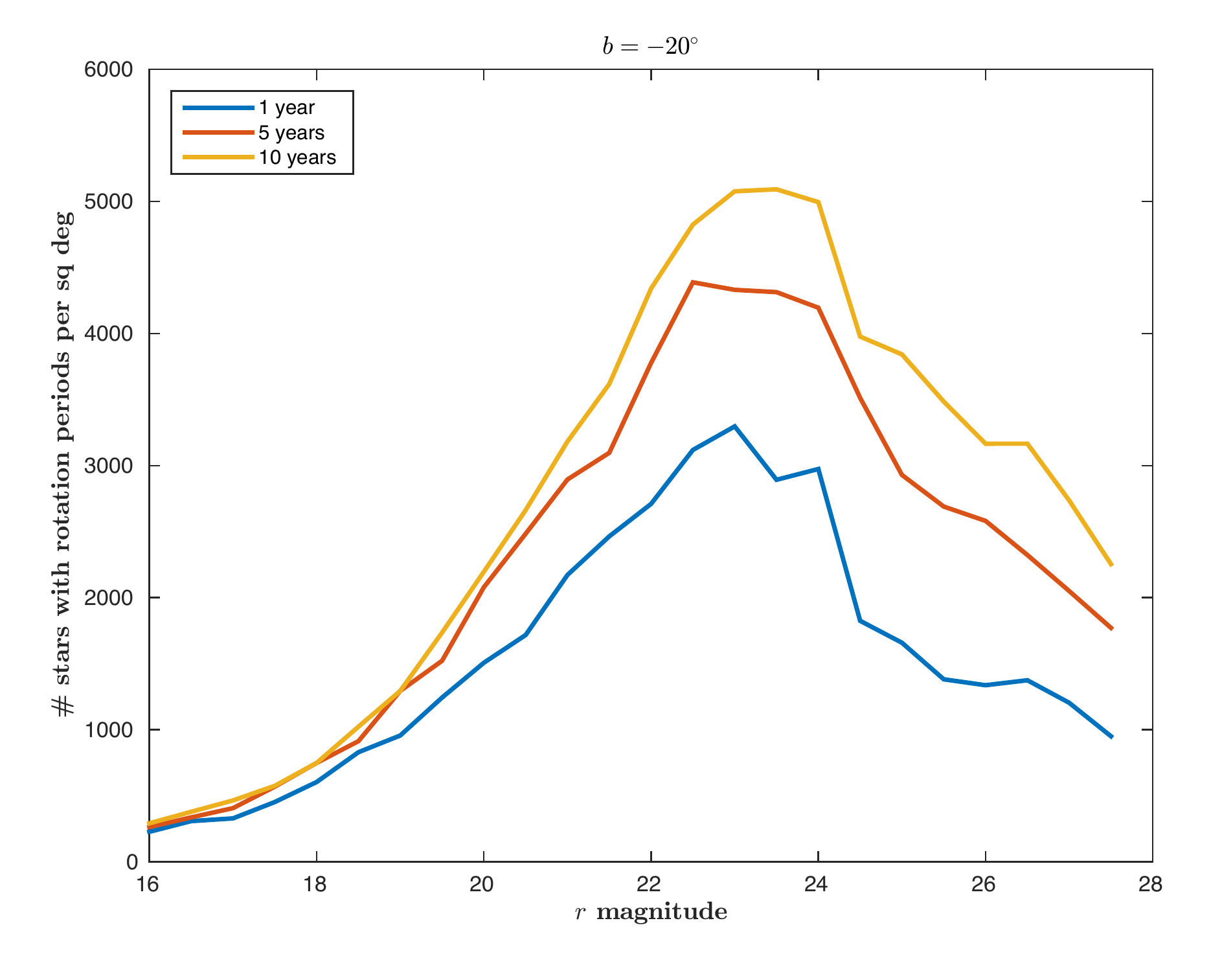}
\includegraphics[width=3in]{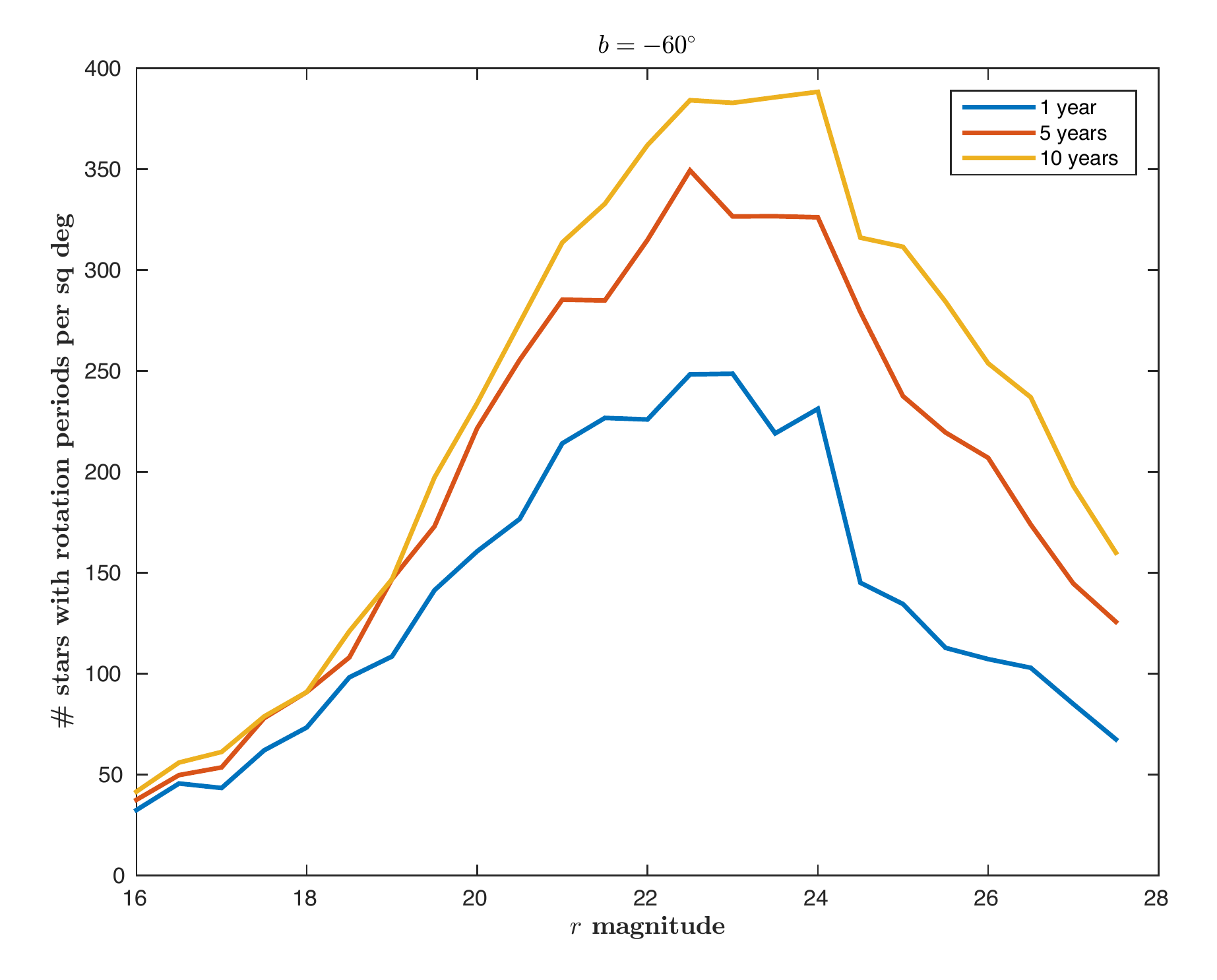}
\caption{
Results of LSST rotation period yield simulations. Left: Density of stars with detected periods at $b = -20$. Right: Density of stars with detected periods at $b = -60$.}
\label{fig:rotsim}
\end{figure}

Finally, we scaled the results from the 20,000 star target sample to a one square degree field, and found the density of stars with detected periods (number per square degree) after 1, 5 and 10 years of LSST observing.  The density distributions as a function of $r$ magnitude are shown for fields at two different  latitudes in Figure \ref{fig:rotsim}.  In each magnitude bin, there are thousands of stars per square degree at low Galactic latitudes, and hundreds per square degree at high latitudes.

\subsection{Flare Simulations}

Forward modeling of flare star light curves allows us to predict the density of detectable flare stars on the sky as a function of limiting magnitude. Estimating this target density is critical for planning systematic follow-up studies of magnetically active stars, as well as for predicting the impact of flares as a contamination source for other variability studies with LSST.

\begin{figure*}[]
\centering
\includegraphics[width=5in]{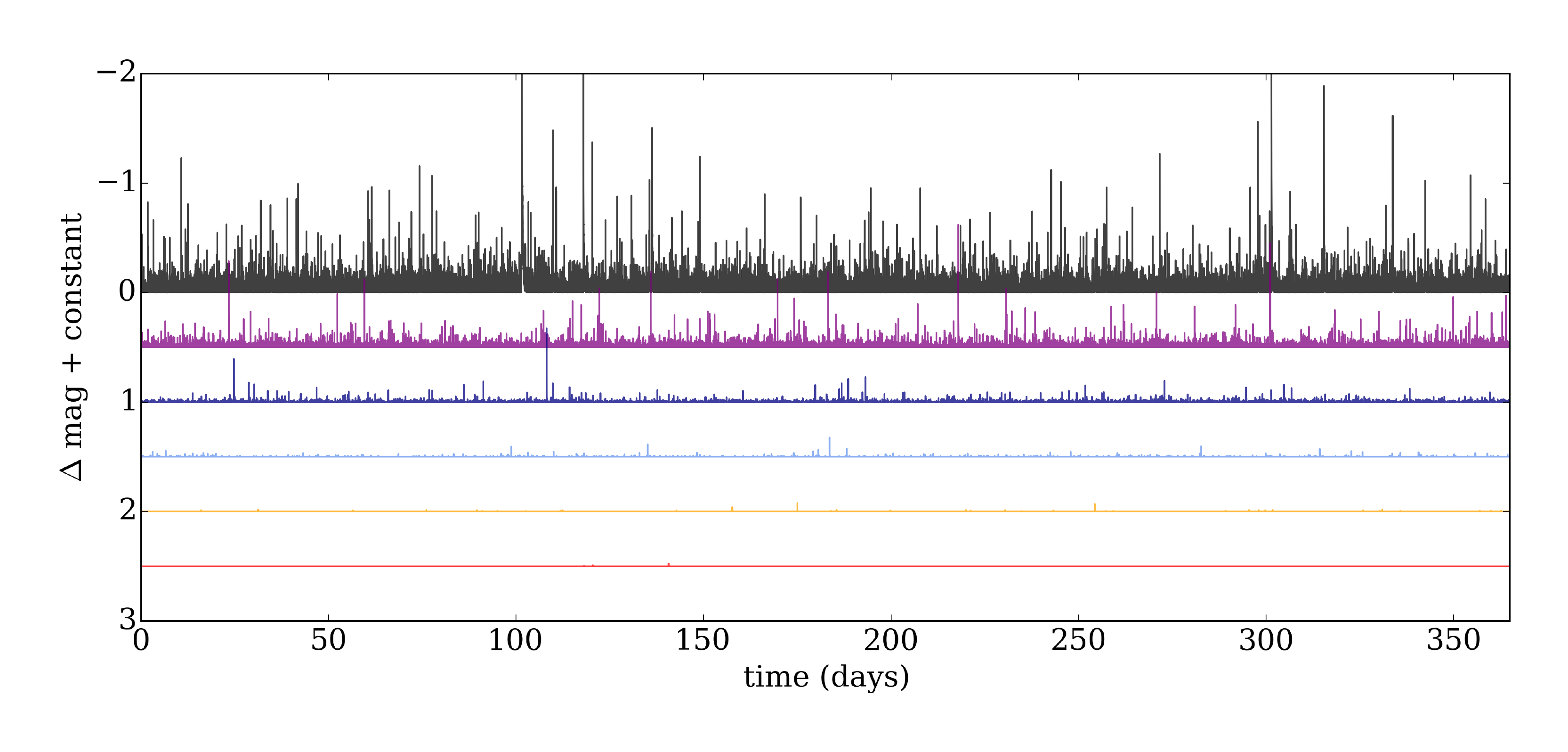}
\caption{
Simulated flare star light curves for six levels of flare activity. Each light curve is 1 year in duration, sampled at a 1 minute cadence. Light curves from top to bottom decrease by an order of magnitude each in their cumulative rate of flares per day, from $\alpha=1$ (black line) to $\alpha=-4$ (red line).
}
\label{fig:flarelc}
\end{figure*}

To determine accurate flare yields for stars at the sparse LSST cadence, we first simulated light curves with a range of flare rates using a Kepler-like cadence of 1 minute.  Individual light curves contained 1 year of continuous simulated flare data, and were repeated 10 times to create a 10-year light curve. The simulations used a typical flare occurrence rate as a function of energy which is described by a power law:

\begin{equation*}
\log \nu = \beta \log \varepsilon + \alpha \;,
\end{equation*}

\noindent where $\nu$ is the cumulative number of flares per day, and $\varepsilon$ is the flare energy normalized by the quiescent luminosity of the star. 
We fixed the slope at $\beta=-2$, a value commonly used for the Sun and nearby flare stars \citep{hawley2014}. The total number of flares is governed by the power law amplitude parameter $\alpha$ which we varied from $\alpha=1$ for very active stars to $\alpha=-4$ for inactive stars \citep{hilton2011}, resulting in the six simulated light curves shown in Figure \ref{fig:flarelc}.  The profile of each flare was computed using an empirical model derived from Kepler flare observations \citet{davenport2014b}. Only classical, single-peaked flares were simulated, though complex flare morphologies resulted for the active star light curves due to serendipitous flare overlaps.

Within each of the four TRILEGAL fields (see \S2.2) we randomly chose 50,000 stars for flare simulation.  The quiescent $g$-band magnitude was used to compute the typical LSST photometric uncertainty $\sigma_g$ for each target. Each star was placed into a bin depending on its temperature and age, using  $T_{eff}$ bins of 1000 K (roughly G, K and M spectral types) and 3 age bins: $<250$ Myr; 250 Myr--1.5 Gyr; and $>1.5$ Gyr.
The age and spectral type determine which of the 6 simulated light curves is used for the simulation. For each temperature bin, the flare rate parameter ($\alpha$) is decreased with age using a simple ad hoc prescription since the detailed evolution of flare rates with stellar age is still unknown.  Indeed, the LSST data will be uniquely able to provide new constraints on the flare rate-age relationship. Although no galactic kinematic information is provided for these stars, it will be important to compare ages we obtain from rotation and flare activity to ages obtained from kinematic population studies.

Once the appropriate light curve is chosen for each star (depending on its temperature and age), it is then sampled over 10 years using the cadence model described in \S2.1.  Only the $u$ and $g$ band visits result in flare detections, since flares are intrinsically blue events.  Thus, each light curve is sampled approximately 100 times over the 10 year period, according to the $minion\_1016$ cadence model (which has many fewer $u$ and $g$ observations than in the redder filters). 

Flares in the sampled light curves are defined as single epoch (bright) outliers.  We used a conservative limit of a measured flux that is 0.1 magnitudes brighter than the median magnitude of the star (in the $u$ and $g$ bands) for a confident flare detection.  This corresponds to $>$10$\sigma$ and also should reduce false detections from other stellar variability e.g. from starspots, which is typically $<$ 0.03 magnitudes.   We note that if the LSST cadence includes two 15-second exposures as part of the baseline observing plan for each visit, then flares can be verified by comparing the magnitudes in the two exposures.  Since flares typically last for several minutes, and large flares last for several hours, the two exposures should both show the flare brightening, which will reduce the number of false positives.  Access to the data from the two exposures as part of the Level 3 LSST data products will therefore be essential for identifying real flares and rejecting noise events.

\begin{figure}[]
\centering
\includegraphics[width=3in]{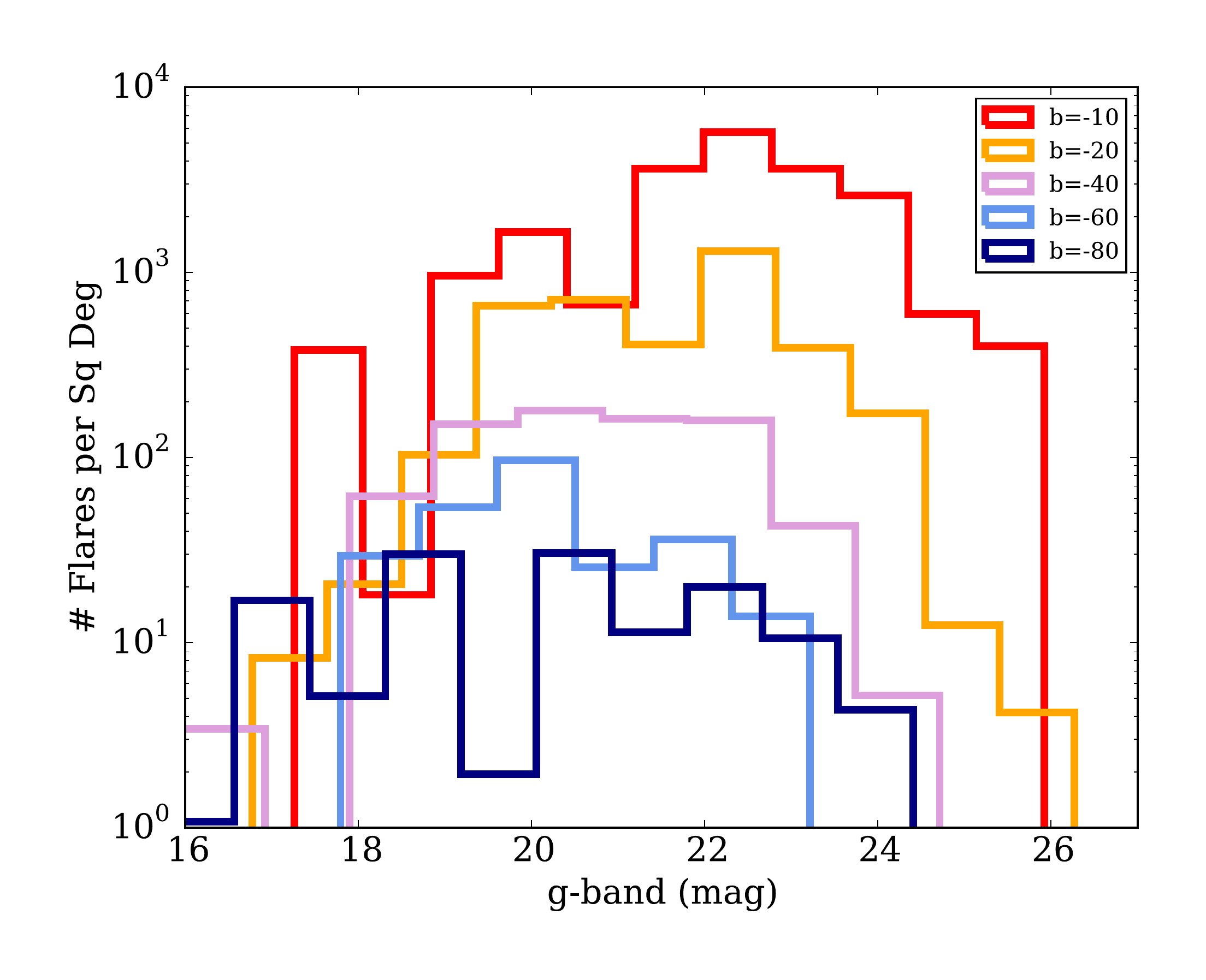}
\includegraphics[width=3in]{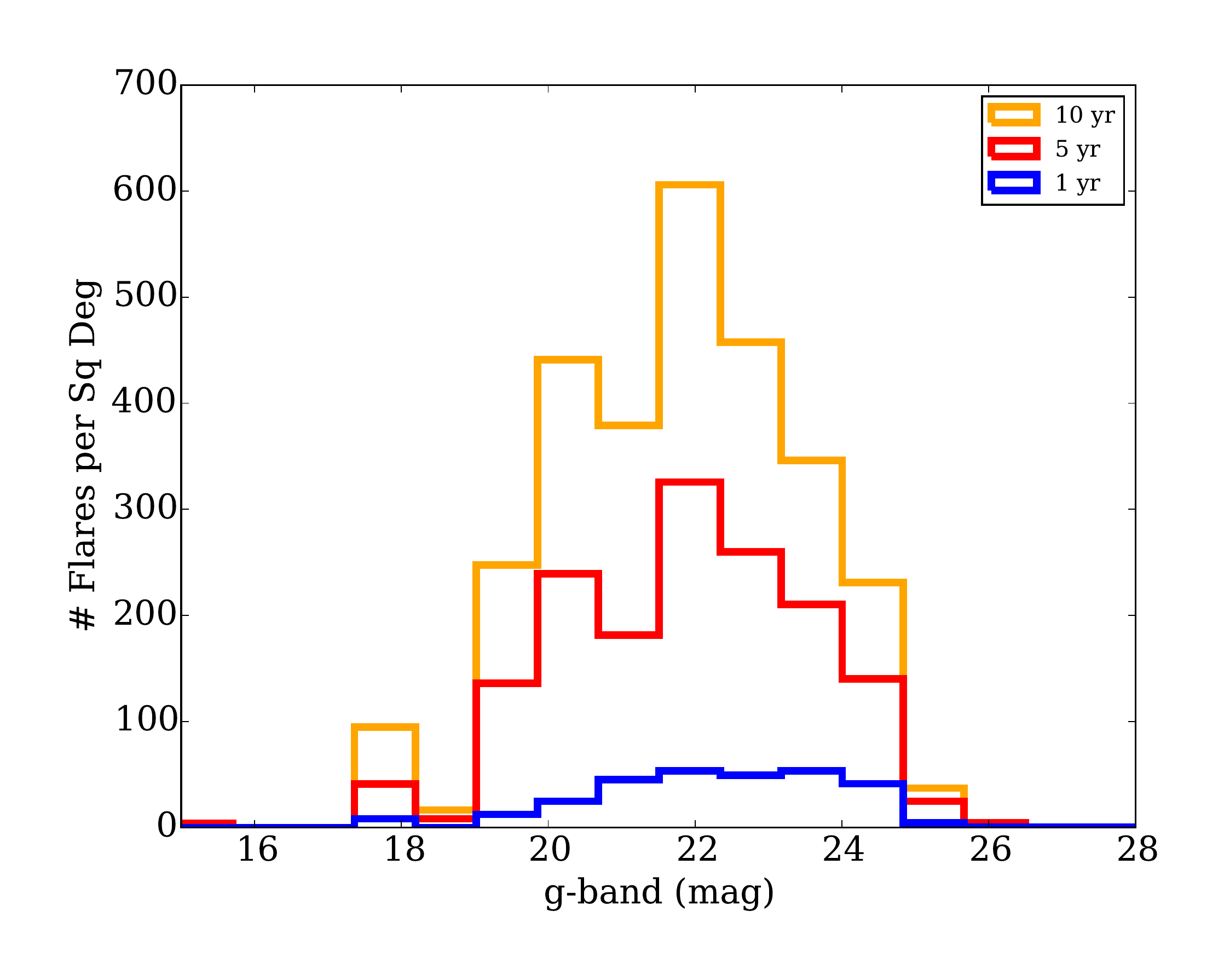}
\caption{
Results of the LSST flare yield simulations. Left: Spatial density of detected flares using 10 years of LSST data versus stellar quiescent apparent magnitude for the four test fields. Right: Flare yield for 1, 5, and 10-year baselines versus apparent magnitude for the $b=-20$ line of sight.}
\label{fig:flaresim}
\end{figure}

For each line of sight we scaled our recovered flare yields from the 50,000 sample stars to a one square degree field. This produced the predicted flare rates for the 10 year LSST survey shown in Figure \ref{fig:flaresim}.
M dwarfs dominate both the total number of stars in each field, and the resulting flare yields, making up more than 97\% of the stars with detected flares.  There are two reasons: In G and K dwarfs the flare rate drops very rapidly with age, and there are not many young stars even in the lowest latitude field that we modeled. Flare rates for these stars will be much better determined with LSST by studying young open clusters.  Also, the flare contrast is reduced in GK stars, so reaching the 0.1 magnitude limit for flare identification requires a very large flare.  If the limit for identifying flares can be reduced as the survey progresses, then more GK flares may be found.

The flare rates as a function of galactic latitude are very close (within $\sim$10\%) to those predicted by \citet[][see also LSST science book]{hilton2011}, who performed a similar analysis based on flare rates for M dwarfs from ground-based data.  The flare rates vary by more than 2 orders of magnitude depending on latitude.  Our results show that LSST should detect approximately 1 flare per square degree per exposure at mid to high galactic latitudes, and 10-100 at lower latitudes.  
 At lower latitudes, not only are there more flare stars per square degree, and larger rates of flares due to younger stars near the Galactic mid-plane, but we find the quiescent brightnesses of flare stars reach to fainter magnitudes and further distances.  For high latitude fields, flare stars are only visible at $< 2$ kpc, while in the lowest latitude fields they may be seen out to $\sim$10 kpc in our simulations. However, the largest simulated flares only reach $\sim$2 magnitudes in peak amplitude (in $g$), and quiescent stars are only simulated down to 28th magnitude, limiting our ability to find the rare but interesting flare transients that result from very bright flares on very faint stars. It is clear from the simulations that these flare transients (i.e. flares from stars whose quiescent luminosity is below the detection threshold) will be primarily found at lower latitudes.

\subsection{Field stars (rotation and flares) followup observations and capabilities}

\begin{figure}
 	\begin{minipage}[c]{0.6\textwidth}
		\includegraphics[width=3in]{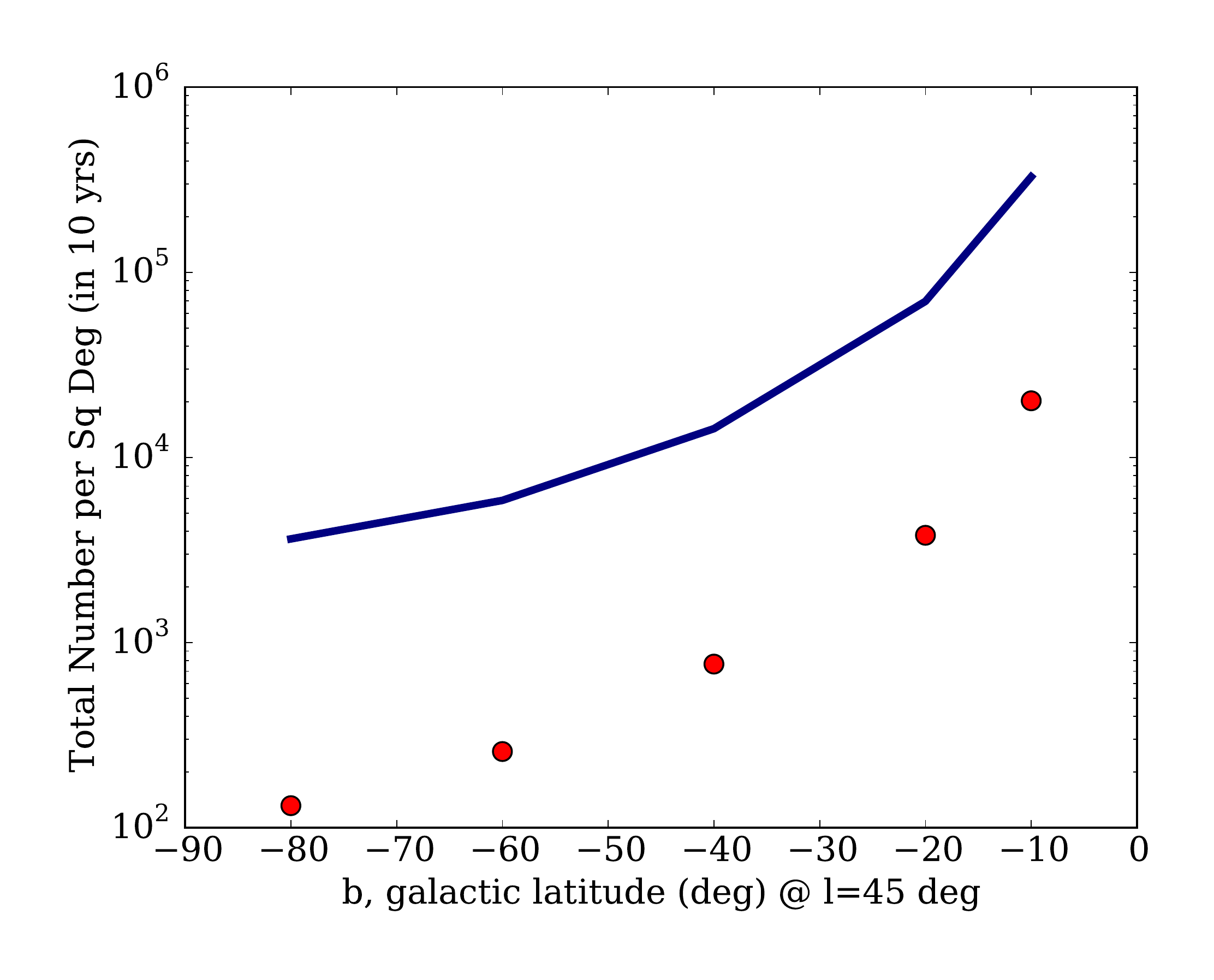}
	\end{minipage}\hfill
  	\begin{minipage}[c]{0.4\textwidth}
 		\caption{Density of stars with detected flares (red circles) and stars with detected rotation periods (blue line) as a function of galactic latitude, for the full 10 year LSST simulation.}
		\label{fig:summaryplot}
  	\end{minipage}
 \end{figure}

Figure \ref{fig:summaryplot} is a summary plot showing the final results of our field star simulations for rotation and flares.  After 10 years of LSST survey observations, the flare stars (nearly all M dwarfs) will number between 100-10$^4$ per square degree, while the stars with measurable rotation periods are almost two orders of magnitude larger, between 10$^4$ and 10$^6$ per square degree (summed over all magnitudes).  The numbers in each category that are found per year increase roughly linearly (see Figures 3 and 5), so in the first year the samples will be about a factor of 10 less.  If stars are restricted to $r$ $<$ 24, the rotation sample is reduced by another factor of 2 or so.  Thus, in round numbers, there will be 10s-100s of stars per square degree to follow up each year, depending on the latitude being observed.

Followup photometry and spectroscopy for the stars with rotation periods will be required in order to
confirm their periods, identify them as bona-fide single main sequence dwarfs, obtain stellar parameters (temperature, gravity, metallicity), 
investigate their magnetic activity properties (e.g. Ca II H and K, H$\alpha$
emission) and thereby produce a sample that will be useful for gyrochronology and age estimation.
We do not expect that it will be necessary to followup every one of these targets.
A thorough characterization of a fraction of them will suffice
to gain an understanding of the population of targets and to calibrate
relations between their spectroscopic properties and their measured (with LSST) $ugrizy$
magnitudes.

Confirming the rotation periods using followup photometry will be necessary 
for two reasons: (1) to measure
rotation periods shorter than a few days which may be missed by LSST; and
(2) to ensure that the variability appearing in LSST light curves is
indeed caused by starspot modulations and is representative of a rotational
signal.
Supplementing LSST data with short-cadence followup photometry
will be particularly necessary for young populations that are likely to have rapidly rotating stars.
By comparing LSST rotation periods with those measured with followup photometry we will
determine a typical success rate, e.g. the fraction of LSST rotation
periods that show good agreement with the rotation periods measured using
followup photometry.
We anticipate that followup photometry of
$\sim$10,000 targets will be required in order to determine meaningful success rates for LSST periods.  We propose that the targets will be distributed as approximately 100 objects in each of 100
different one-degree square fields.  The targets (V=16-24) will span the range of spectral type, period and age of the full gyrochronology sample and will be distributed in RA, though of course concentrated in the southern hemisphere LSST footprint.
Each field should be observed for three visits of about 20 minutes each per night for a month, thus 30 hours per field, or 3000 hours total.  The brightest targets will be accessible with $<$ 3m telescopes, and for example would make good use of a facility like LCOGT.  However, the majority of the targets will be best observed with a 3-5m facility operated in a survey mode, such as ODI on WIYN or DECam on the Blanco telescope.  

Moderate resolution (R$\sim$5000) spectroscopic followup will also be needed.
Spectra are required in order to establish a sample of well-characterized stars for further study (e.g. to determine activity cycles, magnetic fields, etc. see below).
Multiple spectroscopic observations ($\sim$ 3) per star are needed to
determine magnetic variability and binarity.
The spectra will also be used to train a photometric classification system
to characterize a much larger sample of gyrochronology targets.
Spectroscopic $\log g$, effective temperatures and metallicities will be
mapped onto the LSST photometric system using this representative sample of
stars and then applied to the full LSST catalog.
The number of targets required is driven by our desire to sample each of
these physical parameters for stars ranging from late F to M spectral
types.
We anticipate that LSST photometry will enable determination of $T_{eff}$ to
within 200K, and metallicity and $\log g$ to within $\sim$ 0.2 dex.
Given the range of these parameters among our target stars (3000 K, 1.5 dex,
and 1 dex respectively), we would therefore want to populate roughly 1000
bins.  With 10 stars per bin, this implies the need for  $\sim$ 10,000
stars, or (with 3 spectra per star), 30,000 spectra in total.  Ideally there would be significant overlap with the 10,000 stars targeted for photometric followup.

The instrument of choice for the spectroscopic followup would be a MOS with at least 100 fibers, at R=5000, covering a one square degree field.  Again the brighter targets could use such an instrument on a 3-5m telescope (e.g. WIYN/Hydra, DESI/Mayall) while 6m telescopes such as MMT and Magellan also have existing or planned instrumentation that would be suitable.  For the fainter targets, an 8-10m telescope would be required to obtain good S/N ($>$ 20) in reasonable exposure time.  Assuming 1-4 hours of exposure time, 3 exposures per field and 100 fields, this program will require $\sim$ 1000 hours on each facility (3-5m, 8-10m).   Note that our target density is such that we would use only a fraction of a modern MOS with $\sim$ 1000 fibers in each field.  Thus, survey facilities that enable use of fibers for several different programs simultaneously (e.g. such as SDSS) would be well-suited for this program.

Finally, a subset of $\sim$ 100 field stars from the rotation sample will be chosen for more extensive followup after the initial characterization.  This will include high resolution spectroscopy at R=20,000-100,000 to determine radial velocity, rotation velocity (v sin $i$), and metallicity (including especially Lithium abundance).  It is anticipated that these stars will form the activity cycle sample for longterm monitoring of Ca II H and K and photometry, and determination of magnetic field properties as described in \S4 and \S5 below, and the followup capabilities and time needed are included under those sections. 

For the flare star followup, the same type of spectroscopic (R $\sim$ 5000) characterization data are needed as for the rotation sample.  Spectra of approximately 10,000 stars will be used to characterize magnetic activity and binarity, and to provide a sample large enough to calibrate flare rate-age relations.  Kinematic data for population assignment (thin disk, thick disk) will also provide additional information for mapping the individual stars into age bins.  Since these stars are mostly M dwarfs, they will be fainter and require more large telescope resources, thus we anticipate the 2000 hours will be split as 500 hours on 3-5m telescopes, and 1500 hours on 8-10m telescopes.   

A special opportunity for followup of rare ``superflares'' on solar type stars will also be possible from the LSST alert stream.  Such superflares have been identified in Kepler imaging data, but are $>$100 times more energetic than any flares that have been observed on the Sun.  They occur very rarely, perhaps once per 1000 years, but the large sample of G dwarfs being observed with LSST means that there is a chance perhaps a few times per year that such a flare will be observed.  In this case, a single object spectrograph with broad wavelength coverage and moderate resolution should be immediately deployed to obtain spectra for the rest of the night at high time cadence, in order to follow the flare evolution.  Multicolor photometry from an imaging camera is also essential.  The source of the white light continuum emission for these solar-type superflares is unknown, and such a dataset would provide an unprecedented opportunity to examine the physics of this radiation.  Superflares are also important for planetary habitability (including our own!).  Depending on the brightness of the star, the photometric and spectroscopic followup could be obtained with anything from $<$3m to 10m telescopes (though fainter targets are more prevalent, and therefore more likely).  It will be important to verify the flare event by examining both of the 15 second LSST exposures, and to have the LSST transient broker and alert stream operational to direct followup resources within a few minutes of the flare event detection.  We anticipate perhaps 1-3 events per year, requiring 10-30 nights (100-300 hours) of followup observations over the 10 year LSST survey period.

\section{Open Clusters}

Taking advantage of the survey duration, photometric precision, and
sky-coverage of LSST, we can study variability in open clusters over
timescales from days to years.  With 0.005-0.01 mag precision
between $\sim$16-20 mag (griz), studies of stellar rotation, differential
rotation, starspot evolution and magnetic cycle amplitudes and periods
will be possible for cool dwarf stars (spectral types F, G, K, and M) with
ages up to about 2 Gyr.  These stars should also exhibit strong flaring
activity, which may extend even to older clusters for the M dwarf population.
LSST will have access to several hundred Southern clusters, but only a subset
of them are sufficiently rich in stars and at a distance that will allow
us to study their cool dwarfs with LSST.  There are 227 open clusters
in the current LSST footprint that have an
estimated membership count of over 100 stars and distances such that M dwarfs are accessible to LSST (i.e. M0 stars have $V<$20).  We have identified three such
clusters and use them here as a case study to demonstrate the scope of
LSST observations in clusters and the need for followup facilities for
cluster work.

With an age of 170 Myr and a distance of 1200pc, NGC 5316 represents the
early main-sequence phase of cool star evolution.  From literature
studies we estimate that the cluster currently contains 90 known G dwarf
members.  We have fit a \citet{kroupa2001} Initial Mass Function (IMF) to
the G dwarf mass distribution and estimate that the cluster was formed with
$\sim$240 K dwarfs and $\sim$1650 M dwarfs. Most of these $\sim$2000 members
should be accessible from the deep LSST observations over a wide FOV.

NGC 2477 (820 Myr) and IC 4651 (1.7 Gyr) probe the mid-point and the upper
limit of the age-range we can study with LSST.  At a distance of 1450pc
NGC 2477 gives us access to G, K, and M dwarfs 
IC 4651 at 900pc allows us to study older K and M dwarfs (the G dwarfs will be too bright).  We estimate that NGC 2477 and IC 4651 currently contain
85 and 75 G dwarfs, and IMF fits suggest that they formed with $\sim$230
and $\sim$215 K dwarfs and $\sim$1600 and $\sim$ 1500 M dwarfs, respectively.
That is $\sim$1800-1900 cool dwarfs per cluster.

\begin{figure}[]
 	\begin{minipage}[c]{0.6\textwidth}
\includegraphics[width=3in]{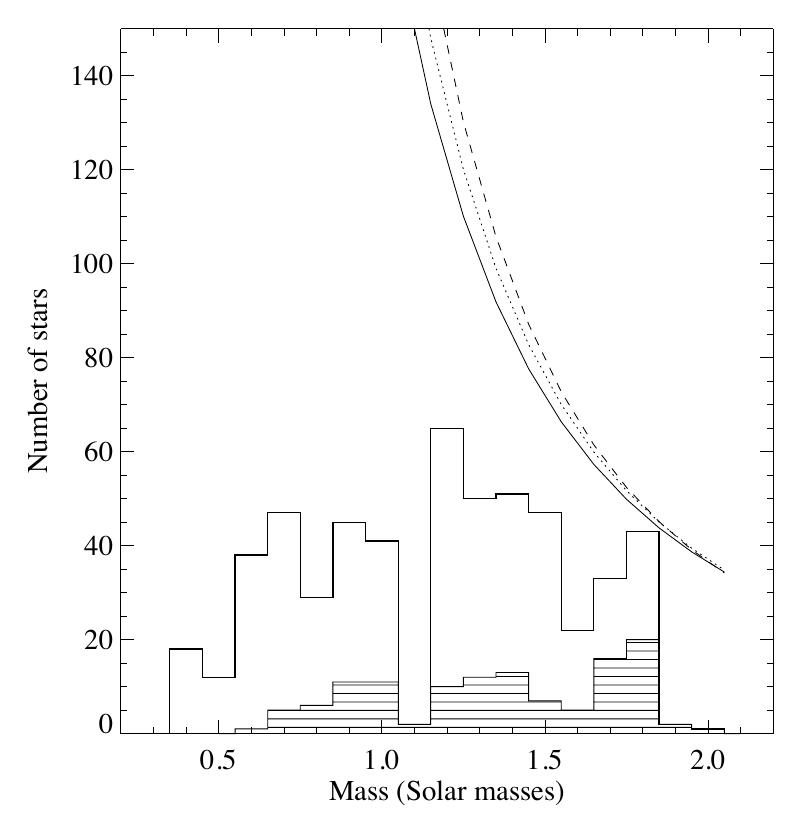}
	\end{minipage}\hfill
  	\begin{minipage}[c]{0.4\textwidth}
\caption{Figure 15 from \citet{meibom2002}, showing the IC 4651 cluster members from radial velocity measurements (shaded histogram), photometry (open histogram) and three possible IMF models (solid, dotted, dashed lines).  The Kroupa model used for our membership estimations is the dashed line. }
\label{fig15}
  	\end{minipage}
\end{figure}

While dynamical evolution (e.g. mass segregation, dynamical evaporation)
is known to remove the majority of low-mass members from the central (core)
region of clusters over a time-scale of $\sim$1 Gyr \citep{naa97,meibom2002},
the large FOV of LSST will easily capture the entire cluster as well as
the foreground and background field population. 
Figure \ref{fig15}, which is a reproduction of Figure 15 from \citet{meibom2002},
shows the currently known members (histograms) and the predicted fits for
three possible IMF models for the relatively old (1.7 Gyr) cluster IC 4651.
These data show only members within the central 10 arcmin of the cluster.
It is clear that membership determination will require significant followup resources such as proper motions, parallaxes,
and radial velocities. In particular, GAIA will provide astrometric data
for stars brighter than V$\sim$20, and LSST itself will produce proper
motion catalogs which may be useful for membership determination.  However,
radial velocities will be needed for at least some fraction of the objects
in order to robustly establish cluster membership and binarity status.

\subsection{Open Clusters followup observations and capabilities}

Similar photometric and spectroscopic (moderate resolution, MOS) followup observations as for the field star sample will be needed for the 227 southern clusters in the LSST footprint, to confirm rotation periods and to obtain accurate stellar parameters and magnetic activity levels.  Assuming 30 hours of photometric monitoring (spread over 1-3 months, depending on cluster age and therefore sampling cadence) for each cluster gives about 7000 hours total.  About 10\% of the clusters have very bright members (solar type stars have V$\sim$12) and the entire cluster population can be captured with $<$3m facilities with wide field imaging capability.  The brighter members of all the clusters (G and K stars) can also be observed in broadband filters with $<$3m telescopes.  Fainter members (M stars to V$\sim$24) and narrowband filter observations will require followup photometry with 3-5m telescopes.  We estimate approximately 3500 hours (half the time) on each size of facility.

Selection of cluster candidate members for followup spectroscopic observations will be
guided by isochrone fits to the color-magnitude diagrams (produced by LSST
and followup photometry).  Stars that fall on the cluster CMD and that have
starspot signatures enabling rotation period measurements and/or exhibit
flares will be the highest priority for followup. Typical contamination by
field stars can be as high as 50\% for G stars, and rise to 80-90\% for K and
M dwarfs depending on the Galactic latitude of the cluster (see e.g. NGC
5316). The spectroscopy can be carried out on 3-5m telescopes with MOS capability, as for the field stars.  Each of the 227 clusters will likely have several thousand candidate members, so the targets can fill a 1000 fiber MOS and each cluster will require about 3 pointings (of 1-4 hours) to get all the candidates.  Again most clusters have relatively bright members (V$<$22) so about 1000 hours on 3-5m telescopes will be sufficient.   The fainter candidates (V=22-24) will need about 500 hours on 8-10m telescopes. 

From these initial followup data, promising candidates can be chosen 
for membership confirmation with high resolution radial velocity observations.  These will also allow identification of binary stars, which is important since the stellar parameters determined from the photometry are affected.  The rotation and activity evolution can also be significantly altered in binary systems.  High resolution (R $>$ 20,000) multi-object spectroscopy on 6-8m telescopes is needed
(e.g. an instrument such as Hecto-echelle on the MMT), and at least 3 spectra per target.  We note that GAIA
will be capable of identifying binary stars only out to 250pc which will exclude
nearly all clusters of interest.  However, the LSST astrometry may prove very useful for establishing kinematic membership.  With 1000 targets per cluster needing the detailed high resolution data, 3 pointings per target to determine binarity, and 1-4h exposures, approximately 2000 hours on 8m telescopes with high resolution 1000-fiber MOS will be required.

\subsection{Detailed cluster evolution sample}
LSST observations of clusters over a range of ages will play an essential
role in studying the evolution of stellar astrophysical phenomena (such as
activity cycles and flare rates) as a function of stellar mass, and in
calibrating relationships between stellar age, rotation, and activity, to
further develop stellar age-dating methods such as gyrochronology.  To this end, we propose that a subsample of 11 open clusters, spanning a range of ages, be observed even more extensively.  If possible, these clusters should be targeted  with LSST deep drilling (or special survey) fields.  If those are not available, then more extensive followup observations will be needed.

The detailed cluster evolution sample comprises clusters that were chosen to lie in or near the LSST
footprint ($\delta < 5$), but more than 5 degrees from the galactic equator
to mitigate source confusion issues. (However, note that NGC 5316 which is discussed above is included in the sample although it has b$\sim$0.)  The clusters were also required to have $V$ between 12.5-16 for solar type stars ($V$ between 16.5-20 for the M0 dwarfs) to ensure that the majority of the lower main sequence
is accessible to LSST with reasonable SNR and without significant CCD
saturation. We further winnowed our list by requiring the clusters to have
well-established metallicities. This process results in 10 Southern clusters. We also include the
canonical cluster M67, which lies slightly north of the LSST survey limit (declination = +12 degrees), but which would be a very worthwhile target for an LSST special survey.
The 11 clusters in the sample are shown from oldest to youngest in Table 1.  They span ages from $\sim$ 5 Gyr to 100 Myr, enabling they study of a wide range of stellar rotational and activity evolution.

\begin{deluxetable}{lcccccccccccc}
\tablecolumns{11}
\tablewidth{0pt}
\tabletypesize{\footnotesize}
\tablecaption{LSST Cluster Evolution Sample.\label{clustertable}}
\tablehead{
	\colhead{Cluster}&
	\colhead{RA}&
	\colhead{Dec} &
	\colhead{$l$} &
	\colhead{$b$} &
	\colhead{$N$} &
	\colhead{Dist} &
	\colhead{$E(B-V)$} &
	\colhead{Age} &
	\colhead{$[Fe/H]$} &
	\colhead{$V_{G2V}$} \\
	\colhead{}&
	\colhead{(hh:mm)}&
	\colhead{(dd:mm)} &
	\colhead{(deg)} &
	\colhead{(deg)} &
	\colhead{members} &
	\colhead{(pc)} &
	\colhead{(mag)} &
	\colhead{(Gyr)} &
	\colhead{} &
	\colhead{(mag)}
	}
\startdata
NGC 6253  & 16:59 & -52:42  & 335.5 &  -6.2 & 255 & 1511 & 0.2 & 5.012 & 0.43 & 16.26 \\
M 67 & 8:51 & 11:48 & 31.9 & 215.7 & 562 & 908 & 0.059 & 3.90 & 0.00 & 14.67 \\
Ruprecht 147  & 19:16 & -16:15 & 21.0 & -12.7 &  170 & 270 & 0.11 & 2.138 & 0.07 & 12.25 \\
NGC 6208  & 16:49 & -53:43 & 333.8 & -5.7 & 176 & 926 & 0.208 & 1.905 & -0.03 & 15.22 \\
IC 4651  & 17:25 & -49:56 & 340.1 & -7.9 & 325 & 888 & 0.121 & 1.778 & -0.128 & 14.87 \\
NGC 2477  & 7:52 & -38:32 & 253.6 & -5.8 & 1209 & 1450 & 0.291 & 0.822 & -0.192 & 16.44 \\
NGC 1901  & 5:18 & -68:26 & 279.0 & -33.6 & 134 & 406 & 0.021 & 0.708 & -0.331 & 12.87 \\
Collinder 350  & 17:48 & 1:21 & 26.8 & 14.7 & 189 & 302 & 0.302 & 0.513 & 0.00 & 13.07 \\
Ruprecht 110  & 14:05 & -67:28 & 310.0 & -5.6 &  182 & 1241 & 0.312 & 0.501 & -0.234 & 16.16 \\
NGC 5316 & 13:54 & -61:52 & 310.2 & 0.1 & 570 & 1208 & 0.312 & 0.17 & 0.045 & 16.11 \\
NGC 6087  & 16:19 & -57:56 & 327.7 & -5.4 & 379 & 889 & 0.271 & 0.089 & -0.01 & 15.32 
\enddata
\end{deluxetable}

The required special (deep drilling) observing cadence for the clusters
is determined primarily by our desire to measure the full range of rotation
periods anticipated in each. Young clusters such as the Pleiades ($\sim$100
Myr) and M34 ($\sim$200 Myr) have stars rotating with periods from a few hours to
$\sim$10-15 days \citep{hbk+10,meibom2009,mms+11}. At the other end of our cluster
age distribution (up to 5 Gyr), rotation periods will range from ~10-100 days
and require the long time-baseline observations provided by the LSST
(Gonzalez, 2016, Newton et al. 2016). Accordingly, the optimal LSST sampling
cadence for clusters would be logarithmic, starting with multiple visits per
night for a week, followed by approximately daily visits for a month, and
decreasing to the native LSST observing cadence of a few visits per week
for the remainder of the project.  If these are not carried out as special survey fields with LSST,
then these observations would need to be performed on followup facilities.  This would be an extension of the imaging followup described above, adding $\sim$1000 hours of additional high cadence imaging on 3-5m telescopes.  LSST will still provide observations at its normal cadence of these clusters for the duration of the 10 year survey.
Additional observations (see \S4) over the ten-year
period will enable the detection of activity cycles in cluster members, as well
as improved estimation of flare rates.

\section{Activity Cycles}

The cadence of LSST is well adapted to the measurement of long-term variations in the magnetic dynamo, which produce stellar magnetic activity cycles in solar-type stars, similar to the 11-year solar cycle.  Using data for field stars, \citet{lockwood2007} determined a relation between cycle amplitude and chromospheric activity level (log R$_{HK}$) measured from the Ca II HK lines), as shown in their their Fig. 7. 
\citet{mamajek2008} used open cluster data to calibrate this chromospheric activity level with age (see their Fig. 6).
 From these relations, we find cycle amplitudes $>$ 10 mmag for stars with ages $<$ 250 Myr, and cycle amplitudes of $\sim$ 5mmag at an age $\sim$ 1 Gyr.  (For reference, the age of the Hyades is approximately $\sim$ 625 Myr while the age of the Pleiades is $\sim$ 100 Myr.)  Thus, depending on the actual long term precision of the LSST photometry, we may be able to measure cycle amplitudes for stars up to about 1 Gyr in age from the survey data alone for relatively bright stars.  Of the three open clusters described in detail above, measuring cycles in solar-type stars in NGC 5316 and NGC 2477 should be possible from LSST survey data, but the older cluster IC 4651 is only predicted to have 2.5 mmag cycle amplitude, and thus cycles would likely not be measurable without additional observations.

To obtain cycle measurements for older clusters (ideally to the age of the Sun, and even more ideally in the cluster M67), additional photometry with very high precision will be needed.  Clusters targeted for LSST deep drilling fields (see above) could provide these data. It is also of interest to correlate flare activity with HK cycles both in clusters and potentially using active stars that are located in the LSST deep drilling fields chosen for other scientific reasons.

In clusters and the field, programs to monitor Ca II HK (spectroscopically) and photometric variations are needed for studies of activity cycles.  Photometric variations in the visible are dominated by cool spots on both short (rotational) and long (cyclical) time scales.  Relatively larger photometric cyclic variations are typically seen in the younger active stars.  Chromospheric variability on rotational and cycle time scales is also evident in the Ca II H and K resonance lines, which are accessible to ground-based observations in the blue-visible.  These features arise from bright magnetic regions on a star.  In young, active stars spots tend to be the dominant concentrations of magnetic flux until about an age of 2 Gyr.  At ages older than about 2 Gyr photometric amplitude variations become more difficult to detect, declining to only 1 mmag at the age of the Sun. Spectroscopic observations of the Ca II lines are then preferred since cycle variations in the strengths of these features are readily detectable.  At all ages and activity levels, observations of photometric variations and spectroscopic observations of Ca II HK variability are complementary since each diagnostic samples distinct and fundamental constituents of stellar magnetism.  Therefore, a program of HK monitoring in parallel with photometric monitoring (both with LSST and with followup imaging) will provide a more complete picture of the evolution of stellar magnetic fields than either facility alone.

For the long-term cycles program, the original Mt. Wilson HK program initiated and carried out by \citet{wilson1978} provides a guide to the minimum observational frequency required. For solar-like cycles with periods $\sim$10 -- 11 years, 2 -- 4 observations per month were adequate.  For shorter cycle periods of 3 -- 5 years, Wilson and his observers obtained data 5 -- 7 times per month for these stars.  In general, it appears that cycle frequency increases with rotation frequency to some power with a steeper dependence for more active stars.  Therefore, we can expect that more rapidly rotating stars will generally (though not always) have shorter magnetic cycle periods.  Moreover, rotational modulation will add ``noise'' to the cycle modulation.  Hence, the need for more frequent observations during a month for high activity stars in order to accurately infer long-term cycle properties.



Note that in the critical dynamo regime of the M dwarf stars there is a paucity of photometric data.  It is in this region of the H-R diagram where solar-like dynamos, which depend on the presence of an interface region between the outer convection zone and the radiative interior for magnetic field amplification, may undergo a transition when the interior becomes fully convective at approximately spectral type M4. The sparse data that have been obtained so far suggest a range of cycle amplitudes in $V$ of approximately 40 mmag -- 100 mmag 
\citep[see][and references therein]{buccino2014}.
Given the large number of active M dwarfs that will be identified in the field from the rotation and flare investigations (see above), LSST data will uniquely enable investigation of cycle properties in field stars for this crucial regime of dynamo operation where interiors transition from partial convection to whole interior convection.  

\subsection{Activity Cycles followup observations and capabilities}

A sample of $\sim$ 100 field stars will be drawn from the field star sample that has already been characterized with followup photometry and moderate resolution spectroscopy.  These stars will be monitored for long-term activity cycle variations in both Ca II H and K and broad band photometry.  They will be an important addition to nearby star samples since they will investigate different populations - older, lower mass and lower metallicity. 
The 11 clusters in the cluster evolution sample will be similarly monitored for cycle variations and will contain perhaps a few thousand stars in each cluster.  The followup observations that are required include moderate resolution spectroscopy (R$\sim$ 5,000) and $ugriz$ + narrow-band Ca HK photometry with filters approximately 1\AA\ wide.  The monitoring observations are needed twice per month in order to characterize stars that have cycle periods of 1-10 years.  If the 100 star field sample is chosen from a single one degree field, and with an hour per observation, this requires 2 hours per month per field x 12 fields (11 clusters plus one field sample) x 12 months $\sim$ 300 hours per year or about 3000 hours total of $ugriz$+HK imaging followup.  To avoid duplicating observations on different telescopes, the sample must be chosen so that the brightest and faintest stars in the observing field are accessible with the same telescope, hence a 3-5m wide field imaging facility such as DECam on the Blanco will be preferred.  

The moderate resolution spectroscopy to determine Ca II HK variations will have the same cadence requirements, but will require 3-5m and 8m facilities depending on the target brightness, as for the initial characterization followup (see \S2.5). We anticipate 200 hours per year of 3-5m MOS and 100 hours per year of 8m MOS for the field plus cluster samples.

High resolution spectroscopy (R=20,000-100,000) will also be needed in order to provide additional characterization including radial velocity, rotation velocity, metallicity (including Lithium) and magnetic field properties as described in the next section.  The clusters will already have these observations, so they are only needed for the 100 star field sample, hence will only require about 10 hours each on 8m and 25m telescopes.

\section{Magnetic Fields}

Readily accessible chromospheric features such as the Ca II resonance lines are radiative proxies of magnetic field-related activity.  A key goal of stellar astrophysics is to obtain direct measurements of magnetic field properties across the H-R diagram.  The direct measurement of magnetic fields in late-type stars has been challenging because of tangled field topologies that yield no net circular polarization or only a residual polarization signal in favorable geometrical circumstances.  An advance occurred when \citet{robinson1980} utilized unpolarized, white-light measurements to directly detect the presence of Zeeman broadening in the wings of magnetically sensitive lines as compared to insensitive (or much less sensitive) lines from the same multiplet.  This approach, which typically requires spectral resolutions of R $\sim$ 100,000 in the visible, yields a measure of magnetic field strength and fractional area coverage as inferred from a schematic representation of the Zeeman triplet splitting pattern.  Extending this method to the infrared offers real advantages in stellar magnetic field measurements given that Zeeman splitting is proportional to wavelength squared, though in practice the gain is proportional to wavelength since natural line (Doppler) widths increase directly with wavelength.  The minimum resolution requirements in the infrared would be R $\sim$ 60,000.

The white-light approach described above has been extended with modern spectropolarimetric approaches that are particularly sensitive to the large-scale field and toroidal flux component. An example is the long-term polarimetric monitoring of the solar-type star $\xi$ Boo A (G8 V) by \citet{morgenthaler2012}.  These investigators find that the large-scale field on $\xi$ Boo A is characterized by an axisymmetric component that is dominated by its toroidal component.  Interestingly, an earlier study of the presumably fully convective, active dwarf M star, V374 Peg, also reveals a magnetic field structure dominated by a strong axisymmetric component in the presence of only weak differential rotation \citep{morin2008b}. Morin et al. note that this finding is in contrast to dynamo theories that require strong antisolar differential rotation in order to sustain a strong axisymmetric field component while only non-axisymmetric geometries would occur in the presence of weak differential rotation.   Clearly, an expansion to a large sample of stars would allow us to explore the broader applicability of this very preliminary finding.  In addition to global field topologies, the magnetic field properties of starspots based on molecular features intrinsic to these cool regions along with Stokes V observations are a new area of spectropolarimetric investigation.   

Doppler imaging relies on high-resolution spectra to detect the distortions in the cores of absorption line profiles as rotation carries thermal inhomogeneities across the line of sight. Doppler imaging is best applied to stars with high projected rotational velocities so that the signature of a cool spot is resolvable in velocity space as its line-of-sight rotational velocity component traverses the absorption core.  These objects also probe dynamo properties in the limit of rapid rotation and, generally, thick convection zones. Through the application of inversion techniques, Doppler imaging yields a mapping of cool spots on the stellar surface as a function of phase.  The primary targets of Doppler imaging tend to exhibit large polar spots that appear to be long lived, relatively stable structures \citep[see reviews by][]{berdyugina2005,donati2009,vidotto2014}.  Instrumental resolutions of at least 40,000 are needed for Doppler imaging.
In addition to Zeeman broadening detection and net circular polarization measurements, Zeeman Doppler Imaging (ZDI) is actively utilized for discerning magnetic field properties in rapidly rotating stars.  This technique combines Doppler imaging with Zeeman spectropolarimetry to detect rotationally modulated Zeeman components in a magnetically sensitive line.  The resolution requirements are essentially the same as for Zeeman broadening techniques, i.e., R $\sim$ 60,000.  Through this approach the poloidal and toroidal components of the large-scale stellar magnetic field can be deduced to unveil the nature of field topologies in active stars.  The broad conclusions thus far from ZDI and other magnetic field studies suggest that stars more massive than 0.5 Msun and with Rossby numbers near unity are characterized by a mainly non-axisymmetric poloidal component and a significant toroidal component.  In contrast, less massive, active stars seem to produce strong large-scale poloidal and axisymmetric fields \citep[see][]{donati2009}.

Another accessible spectral feature that is uniquely powerful as a radiative proxy for surface magnetic field regions in solar-type stars is the He I triplet line at 1083 nm.  This feature (as well as the weaker He I triplet line at 587.6 nm) appears in absorption in active (plage) regions on the Sun and, by implication, in the magnetic regions on Sun-like stars.  Since these features are relatively weak high-resolution spectroscopy is needed to measure their equivalent widths while also disentangling the contamination due to terrestrial water vapor.  Typically, resolutions in the range of R $\sim$ 50,000 are desirable.  The 1083 nm absorption line is not seen, or appears only very weakly, in the quiet solar (or late-type dwarf stellar) photosphere. The 1083 nm line is spatially correlated with significant concentrations of magnetic flux in the photosphere and sites of X-ray emission in the corona but is otherwise only weakly present in the quiet photosphere.  Thus, the absorption equivalent width of the 1083 nm line in the integrated spectrum of the Sun or that of a solar-type star can yield direct information on the filling factor of magnetic regions, when properly calibrated by models or through empirical approaches \citep{andretta1995}

The next step in this fundamental area of the solar-stellar connection is to obtain (1) long-term polarimetric studies of the cycle modulation of field topologies for F to M dwarfs and (2) extend magnetic field studies to higher sensitivities in order to obtain information on more nearly solar-like stars with solar rotation periods.  The latter will be particularly important in order to understand the extent to which solar global topologies are shared among solar-type stars of different activity levels.

\subsection{Magnetic Fields followup observations and capabilities}

Spectra with resolutions of R $\sim$ 100,000 are required in the visible and R $\sim$ 60,000 in the infrared (where Zeeman splitting is greater for a given field strength).  This will require a high resolution spectrograph on 8m-25m telescopes.  We anticipate defining representative subsamples in the 12 fields that comprise the activity cycle sample (11 clusters plus one field star field) for intensive magnetic field followup observations.  One hour-long observation per month for each field gives $\sim$ 300 hours per year, split as 200 hours per year on an 8m telescope and 100 hours per year on a 25m telescope.  These observations will directly determine how globally averaged field strengths and area coverages change over the course of an activity cycle and will quantitatively establish the correlation between radiative proxies of magnetic activity, such as Ca II H \& K, and magnetic field properties.  

Polarimetric (Stokes QUV) observations (e.g. with an instrument similar to CFHT ESPaDOns) will be essential to establish the field topology.  This will likely require single object observations of about 4 hours at high resolution so will only be feasible for a small sample, $\sim$ 10 stars in each cluster or about 400 hours total split between 8 and 25m telescopes.  The field topologies may also exhibit cycle-dependent properties, however monitoring would be extremely time-intensive, requiring several nights (50 hours) per year or 500 hours for the ten year survey, for each star.  These observations would likely be better carried out on nearby brighter field stars so we have not accounted for them in the time-needed table.

\section{Summary}

The followup resources and time needed to carry out this particular science topic on rotation and magnetic activity in low mass stars are listed in the following tables.  High priority resources include wide field (one square degree) imaging with both broad band ($ugriz$) and narrow band (Ca II H and K) filters and highly multiplexed (1000 fiber) spectrographs at moderate (R=5000) and high (R=20,000-100,000) resolution.  Polarimetry is also extremely valuable for magnetic field measurements and requires a high resolution spectrograph that has been designed for polarimetric observations.  A range of telescopes from $<$ 3m to 25-30m  will be needed, with the smaller telescopes mainly used for imaging and the larger ones for spectroscopy.  To implement the entire (ambitious!) program outlined here will require significant resources over the ten year period of LSST operations:  4000 hours on $<$3m telescopes; 11,000 hours on 3-5m telescopes; 6500 hours on 8-10m telescopes; and 1200 hours on 25-30m telescopes.

We noted two special infrastructure requests, first in \S2.4 that the individual 15 second images be made available as Level 3 data products (the same as the standard LSST data) and second in \S3.2 that several LSST special survey (``deep drilling'') fields be focused on specific open clusters, including especially the iconic solar age cluster M67 which lies just outside the nominal LSST footprint.

After carrying out this exercise in designing a program to followup on LSST photometry of magnetically-active low-mass stars, we are truly excited about the new frontiers that LSST will open in the investigation of  gyrochronology, flares, activity cycles and magnetic fields, in both open clusters and the Galactic field.  The results of this program will not only advance stellar astrophysics in a transformative way but they will synergistically inform other fields such as the nature and evolution of exoplanet system environments and the evolution of the Galaxy.  We look forward to pursuing this investigation for real starting in 2022!


\end{document}